\RequirePackage[switch, columnwise, running, mathlines, displaymath, 
mathlines]{lineno}
\RequirePackage{docswitch}
\def\flag{apj}
\setjournal{\flag}

\documentclass[\docopts]{\docclass}

\newcommand*\patchAmsMathEnvironmentForLineno[1]{%
  \expandafter\let\csname old#1\expandafter\endcsname\csname #1\endcsname
  \expandafter\let\csname oldend#1\expandafter\endcsname\csname end#1\endcsname
  \renewenvironment{#1}%
     {\linenomath\csname old#1\endcsname}%
     {\csname oldend#1\endcsname\endlinenomath}}%
\newcommand*\patchBothAmsMathEnvironmentsForLineno[1]{%
  \patchAmsMathEnvironmentForLineno{#1}%
  \patchAmsMathEnvironmentForLineno{#1*}}%
\AtBeginDocument{%
\patchBothAmsMathEnvironmentsForLineno{equation}%
\patchBothAmsMathEnvironmentsForLineno{align}%
\patchBothAmsMathEnvironmentsForLineno{flalign}%
\patchBothAmsMathEnvironmentsForLineno{alignat}%
\patchBothAmsMathEnvironmentsForLineno{gather}%
\patchBothAmsMathEnvironmentsForLineno{multline}%
}

\usepackage{lsstdesc_macros}

\usepackage{xspace}
\usepackage{enumitem}
\usepackage{graphicx}
\usepackage{wasysym}
\graphicspath{{./}{./figures/}{.logos}}
\newcommand{\qp}{\texttt{qp}}
\newcommand{\pz}{photo-$z$ PDF}

\newcommand{\mgdata}{bright\xspace}
\newcommand{\Mgdata}{Bright\xspace}
\newcommand{\ssdata}{faint\xspace}
\newcommand{\Ssdata}{Faint\xspace}

\begin{document}

\title{ Approximating photo-$z$ PDFs for large surveys }

\begin{abstract}

Modern galaxy surveys produce redshift probability density functions (PDFs) in 
addition to traditional photometric redshift (photo-$z$) point estimates.
However, the storage of photo-$z$ PDFs may present a challenge with 
increasingly large catalogs, as we face a trade-off between the accuracy of 
subsequent science measurements and the limitation of finite storage resources.
This paper presents \texttt{qp}, a Python package for manipulating 
parametrizations of 1-dimensional PDFs, as suitable for photo-$z$ PDF 
compression.
We use \texttt{qp} to investigate the performance of three simple PDF storage 
formats (quantiles, samples, and step functions) as a function of the number of 
stored parameters on two realistic mock datasets, representative of upcoming 
surveys with different data qualities.
We propose some best practices for choosing a photo-$z$ PDF approximation 
scheme and demonstrate the approach on a science case using performance metrics 
on both ensembles of individual photo-$z$ PDFs and an estimator of the overall 
redshift distribution function.
We show that both the properties of the set of PDFs we wish to approximate and 
the fidelity metric(s) chosen affect the optimal parametrization.
Additionally, we find that quantiles and samples outperform step functions, and 
we encourage further consideration of these formats for PDF approximation.

\end{abstract}

\dockeys{methods: statistical, methods: miscellaneous, astronomical databases: 
miscellaneous, catalogs, galaxies: distances and redshifts}

\maketitlepost

\section{Introduction}
\label{sec:intro}

Upcoming wide-field imaging surveys such as Large Synoptic Survey Telescope 
(LSST)\footnote{\url{https://www.lsst.org/}}\citep{ivezic_lsst:_2008} will 
observe tens of billions of galaxies photometrically, without follow-up 
spectroscopy.
Over the past decade, the Kilo-Degree 
Survey\footnote{\url{http://kids.strw.leidenuniv.nl/}}, Hyper Suprime-Cam 
Subaru Strategic Program\footnote{\url{http://hsc.mtk.nao.ac.jp/ssp/}}, and 
Dark Energy Survey\footnote{\url{https://www.darkenergysurvey.org/}} have paved 
the way for LSST via similar survey strategies on tens of millions of galaxies.
Studies of precision cosmology and galaxy evolution with the anticipated data 
will thus rely almost exclusively on the method of photometric redshift 
(photo-$z$) estimation.
Photo-$z$s are subject to a number of systematic errors, some caused by the 
estimation procedures and others intrinsic to the data itself.
For the purpose of producing public photo-$z$ catalogs, the redshift estimation 
community has thus come to favor methods that provide a photo-$z$ probability 
density function (PDF) conveying the potential for such systematic errors for 
each galaxy in the survey \citep{tanaka_photometric_2018, de_jong_third_2017, 
sheldon_photometric_2012}.

Given that the \pz\ catalogs of ongoing surveys already include $\sim10^{7}$ 
galaxies, and that those of upcoming surveys will include $\sim10^{10}$ 
galaxies, storage of these PDFs must balance accuracy of the catalog against 
limited storage resources.
For example, the LSST's anticipated public catalog will be limited to $\sim100$ 
floating point numbers per galaxy for all information characterizing the 
redshift \citep[Section 4.2.2]{juric_lsst_2017}, including \pz s derived by 
multiple methods.
Furthermore, the problem of storing PDFs is not unique to galaxy surveys.
Gaia\footnote{\url{https://www.gaia-eso.eu/}}, for example, has committed to 
providing a catalog of PDFs of stellar properties including velocities, so an 
approach to optimizing the choice of PDF storage parametrization could benefit 
astronomy more broadly.

\citet{carrasco_kind_sparse_2014} first addressed the question of approximating 
\pz s in the context of a particular galaxy survey and metrics applicable to 
deterministic, not probabilistic, data products.
However, we expect the optimal choice of \pz\ storage approximation to depend 
on the intended science applications and their requirements on \pz\ accuracy as 
well as the properties of the anticipated \pz s.
Different science cases will need different metrics, and different formats may 
be appropriate for different datasets.
In this paper, we address the question of \textit{how} these choices should be 
made, by providing the publicly available \qp\ Python 
package\footnote{\url{https://github.com/aimalz/qp}} \citep{malz_qp_2017} 
enabling each survey to optimize their \pz\ approximation via mathematically 
motivated and science-driven metrics.
We demonstrate this approach on two sets of realistic mock data in the context 
of LSST.

In Section~\ref{sec:methods}, we outline how \qp\ can be used to optimize the 
choice of \pz\ parametrization.
In Section~\ref{sec:data}, we describe the mock datasets on which we 
demonstrate such an analysis.
We present the results of this procedure in Section~\ref{sec:results} and make 
recommendations for the use of \qp\ by the photo-$z$ community in 
Section~\ref{sec:conclusions}.

\section{Methods}
\label{sec:methods}

We have developed the \qp\ Python package to facilitate the approximation of 
one-dimensional PDFs, including \pz s, and comparisons between approximations.

A \texttt{qp.PDF} object is associated with sets of parameters for each 
approximation considered.
Conversions between approximations are facilitated by the 
\texttt{numpy}\footnote{\url{http://www.numpy.org/}} \citep{walt_numpy_2011}, 
\texttt{scipy}\footnote{\url{https://www.scipy.org/}} 
\citep{jones_scipy:_2001}, and 
\texttt{scikit-learn}\footnote{\url{http://scikit-learn.org}} 
\citep{pedregosa_scikit-learn:_2011} tools.
The currently supported parametrizations are described in 
Section~\ref{sec:approx}.

The \qp\ package also provides a few built-in metrics for the accuracy of a 
representation of a PDF relative to a given reference representation.
Built-in plots are made using 
\texttt{matplotlib}\footnote{\url{https://matplotlib.org/}} 
\citep{hunter_matplotlib:_2007}.
A subset of the included metrics is described in Section~\ref{sec:metric}.

Catalog-level manipulations are performed using the \texttt{qp.Ensemble} class 
that serves as a wrapper for operations over collections of \texttt{qp.PDF} 
objects.
Parallelization is facilitated by the 
\texttt{pathos}\footnote{\noindent\url{http://trac.mystic.cacr.caltech.edu/proje
ct/pathos/wiki.html}} \citep{mckerns_building_2012, mckerns_pathos:_2010} 
package.

\subsection{Approximation Methods}
\label{sec:approx}

First, we establish a vocabulary for the approximations.
Each \textit{parametrization} of a \pz\ is defined in terms of the 
\textit{parameters} $\vec{c}$ unique to its galaxy, the \textit{metaparameters} 
$\vec{C}$ shared over many galaxies, and the \textit{format} function 
$\mathcal{F}$ that reconstructs a PDF from its parameters and metaparameters.
A parametrization in turn corresponds to a \textit{representation}
\begin{align}
  \label{eq:definition}
  \hat{p}^{\mathcal{F}, \vec{C}, \vec{c}}(z) &\equiv \mathcal{F}_{\vec{C}}(z; 
\vec{c})
\end{align}
of the approximated \pz, denoted as $\hat{p}(z)$ for brevity.
The \textit{dimensionality} of $\vec{c}$ is the number $N_{f}$ of stored 
parameters per \pz, which are presumed to be scalar numbers unless otherwise 
specified.
The number of elements of $\vec{C}$ is of little significance so long as the 
metaparameters do not have storage requirements that are competitive with those 
of the ensemble of per-galaxy parameters.

While it is possible to construct a catalog where $N_{f}$ is not shared among 
all members but is instead optimized for each galaxy, we restrict this study to 
metaparameters shared over all galaxies considered.
A non-uniform $\vec{C}$ scheme would have the advantage of re-allocating 
storage resources from galaxies whose \pz s can be fully characterized by very 
few numbers to galaxies with highly featured \pz s that require more numbers 
for complete characterization; consider the case of a delta function \pz\ for 
which no more information is preserved for any elements of $\vec{c}$ than the 
location of the peak as opposed to the case of a highly multimodal \pz\ whose 
shape cannot be easily parametrized by any known functional form and would 
benefit from the $N_{f}-1$ parameters wasted on the galaxy with a trivial shape.
However, no galaxy survey mission has yet proposed this scheme, so we could 
thus only speculate as to the potential scope and specific goals of such an 
optimization procedure.
Thus, we postpone such an investigation and eagerly anticipate future 
consideration of this possibility.

\qp\ currently supports conversion of \pz\ approximations between five formats: 
step functions, samples, quantiles, evaluations, and functional forms, which 
may be general mixture models of PDFs from a comprehensive library of those 
with functional forms implemented as \texttt{scipy.rv\_continuous} objects.
When the format function is not specified by the approximation, we refer to a 
generic interpolator function $F_{\vec{C}'}(z; \vec{c}, \mathcal{F}_{\vec{C}})$ 
with its own metaparameters $\vec{C}'$, which must be chosen by the researcher.
\qp\ supports numerous interpolation schemes, including several from the 
popular \texttt{scipy.interpolate} library.

\begin{figure}
  \begin{center}
    \includegraphics[width=\columnwidth]{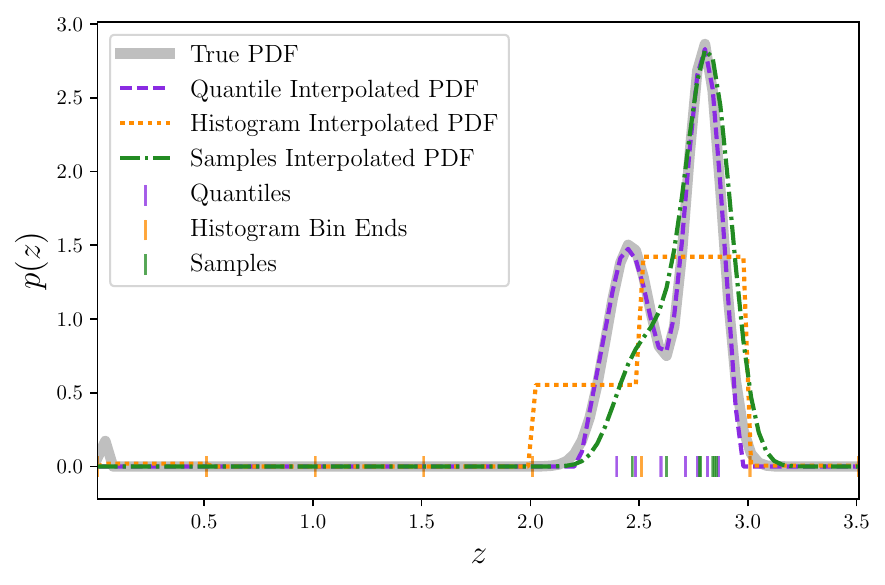}
    \caption{\qp\ approximation of a continuous 1-dimensional PDF (thick, solid 
gray line) using: the step function (orange dotted line), samples (green 
dash-dotted line), and quantile formats (purple dashed line) with the same 
number of stored parameters ($N_{f}=7$ in this case).
    \label{fig:qp}}
  \end{center}
\end{figure}

In this work we consider special cases of three of these formats as candidates 
for large survey \pz\ catalog storage: regular binning 
(Section~\ref{sec:bins}), random samples (Section~\ref{sec:samples}), and 
regular quantiles (Section~\ref{sec:quantiles}).
Step functions and samples have been used in published analyses, and we 
introduce the quantile format because of its favorable statistical properties.
We do not consider the function evaluation format because it is statistically 
very similar to step functions, and we exclude at this time the mixture model 
format because it is only appropriate when the underlying PDFs are actually 
mixture models, which we do not guarantee in this study.
The three formats we investigate are illustrated on a multimodal PDF in 
Figure~\ref{fig:qp}.

In spite of its impressive compression properties, we have not yet included the 
\texttt{SparsePz}\footnote{\url{https://github.com/mgckind/SparsePz}} sparse 
basis representation of \citet{carrasco_kind_sparse_2014}, in which the 
parameters are the integer identifiers of $N_{f}$ mixture model components from 
a library of $\sim10^{4}$ functions.
We omit this format because decomposition with \texttt{SparsePZ} does not 
enforce the condition that the representation be a PDF in the mathematical 
sense of nonnegativity and integration to unity.
While normalizing the integral of a positive semidefinite function is always 
possible (if the endpoints of integration are specified), one can motivate 
multiple schemes for enforcing nonnegativity that result in different 
reconstructions $\hat{p}(z)$.
We postpone to future work the exploration of adaptations of non-positive 
semidefinite representations and inclusion of the sparse basis representation 
in \qp.

For each format, we address the following questions:

\begin{itemize}[partopsep=0pt, parsep=0pt, itemsep=5pt]
  \item When/where has the format appeared as a published catalog format, 
native \pz\ code output format, and science application input format?
  \item What exactly is stored under the format, per galaxy (the parameters) 
and per catalog (the metaparameters)?
  \item What are the a priori strengths and weaknesses of the format?
\end{itemize}

\subsubsection{Regular Binning}
\label{sec:bins}

By far the most popular format for approximating and storing \pz s is the 
piecewise constant step function, also called a histogram binning.
It is the native output of a number of \pz\ codes 
\citep{carrasco_kind_somz:_2014, sadeh_annz2:_2016, cavuoti_metaphor:_2017} and 
the only format that has been used for public release of \pz\ catalogs 
\citep{sheldon_photometric_2012, tanaka_photometric_2018, de_jong_third_2017}.

The metaparameters of the binned parametrization are the ordered list of 
redshifts $\vec{C} = (z_{1}, z_{2}, \dots, z_{N_{f}}, z_{N_{f}+1})$ serving as 
bin endpoints shared by all galaxies in the catalog, each adjacent pair of 
which is associated with a parameter $c_{i} = \int_{C_{i}}^{C_{i+1}} p(z) dz / 
(C_{i+1} - C_{i})$.
The \qp\ histogram format assumes $p(z)=0$ when $z<C_{1}$ or $z\geq 
C_{N_{f}+1}$ and enforces the normalization condition\footnote{
Note that this is not generally equivalent to the heuristic normalization 
condition $\sum_{i} c_{i} = 1$ commonly enforced in public catalogs, such as 
the Sloan Digital Sky Survey's Data Release 8 \citep{sheldon_photometric_2012}.
Unless redshift is treated as a discrete variable, the oversimplified 
normalization condition only holds if $C_{N_{f}+1} - C_{1} = N_{f} (C_{i+1} - 
C_{i})$ for all $i$ (under a regular binning).
}
$\sum_{i} c_{i} (C_{i+1} - C_{i}) = 1$.
The histogram format function $\mathcal{F}^{h}$ is thus the sum of a set of 
$N_{f}$ step functions, making the reconstructed estimator of the \pz
\begin{align}
  \label{eq:binned}
  \hat{p}^{h}(z) &= \sum_{i=1}^{N_{f}}\ c_{i}\ H(z - C_{i})\ H(C_{i+1} - z)
\end{align}
in terms of the Heaviside step function $H$ as the interpolating function $F$.
Though \qp\ supports arbitrary bin ends, here we only consider a regular 
binning, with $C_{i+1} = C_{i} + \delta$ for a constant $\delta = (C_{N_{f}+1} 
- C_{1}) / N_{f}$, as no irregular binning has yet been used for a public 
catalog of \pz s.

In terms of performance as a \pz\ storage format, we should anticipate the 
regular histogram format to be wasteful in terms of information content; a \pz\ 
with a very broad or compact PDF may have many parameters taking the same value 
$c_{i} \approx (C_{N_{f}+1} - C_{1}) / \delta$ or $c_{i} \approx 0$, for broad 
and compact PDFs respectively, that are redundant in storage.
Additionally, we should expect the fidelity of $\hat{p}^{h}(z)$ to depend 
strongly on the bin widths relative to the sizes of and distances between 
features in the \pz s.

\subsubsection{Random Samples}
\label{sec:samples}

Samples are often the native output format of machine learning algorithms due 
to the discrete nature of training sets \citep{de_vicente_dnf_2016}.
Such approaches by default typically produce large numbers of samples, far more 
than can realistically be stored by any survey, so are commonly compressed by 
subsampling \citep{hoyle_dark_2017}.
Samples are easy to use in standard science applications developed for redshift 
point estimates, so they have an established presence in the literature \citep 
{bonnett_redshift_2016}.
The samples format of PDF storage appears elsewhere in astronomy, including the 
Gaia-ESO Survey's commitment to provide multi-dimensional PDFs of stellar 
parameters in the samples format \citep{bailer-jones_gaia_2013}.

The parameters of the samples format are the $N_{f}$ samples $\vec{c}=(z_{1}, 
z_{2}, \dots, z_{N_{f}-1}, z_{N_{f}})$, where $C=N_{f}$ is an implicit 
metaparameter.
The format function $\mathcal{F}^{s}$ that turns samples into a representation 
of the \pz\ is simply the interpolator $F$.
In the tests presented here, we use the Gaussian kernel density estimate (KDE) 
of \texttt{scipy.stats.gaussian\_kde}.
The samples representation is then
\begin{align}
  \label{eq:sampled}
  \hat{p}^{s}(z) &= \mathrm{KDE}_{C'}(z; \vec{c}).
\end{align}

Though samples are an obvious choice for \pz s with narrow features of high 
amplitude, we expect that using a small number of samples from a broad \pz\ may 
increase the variance of any ensemble metrics, as the sampling introduces 
additional shot noise.
The researcher must also choose an interpolation method to reconstruct a \pz\ 
from samples.

\subsubsection{Regular Quantiles}
\label{sec:quantiles}

One parametrization that has not previously been investigated in the context of 
photometric redshifts is that of quantiles, though they have appeared elsewhere 
in the astronomy literature \citep{sun_star_2015, pizzocaro_results_2016, 
laycock_x-ray_2017}.
The quantiles are defined in terms of the cumulative distribution function 
(CDF), which is the antiderivative of the PDF.

Under the quantile format, a \pz\ catalog shares $N_{f}$ ordered CDFs $\vec{C} 
= (q_{1}, q_{2}, \dots, q_{N_{f}-1}, q_{N_{f}})$ where $0 < q_{i} < 1$ for all 
$i$.
Each galaxy's catalog entry is the vector of redshifts $\vec{c} = (z_{1}, 
z_{2}, \dots, z_{N_{f}-1}, z_{N_{f}})$ satisfying $\mathrm{CDF}(c_{i}) = 
C_{i}$, so the quantile format function $\mathcal{F}^{q}$ is the derivative of 
an interpolation $F$ of the CDF.
As with the samples representation, an interpolation function $F$ must be 
chosen for reconstructing the \pz\ from the stored parameters.
\qp\ includes support for numerous \texttt{scipy.interpolate} options as well 
as straightforward linear interpolation.

Our interpolator $F$ in the tests presented here is the derivative of a spline 
function at $z_{1} \leq z \leq z_{N_{f}}$ and linear extrapolation subject to 
consistency with the definition of the CDF.
The quantile representation implemented in this paper is thus
\begin{align}
  \label{eq:quantiles}
  \hat{p}^{q}(z) &=
  \left\{
  \begin{tabular}{cc}
  $\frac{d}{dz} \left[F(z; \vec{c}
  )\right]$ & $c_{1} \leq z \leq c_{N_{f}}$ \\
  $\hat{p}^{q}(c_{1})\left(\frac{\hat{p}^{q}(c_{1})}{2C_{1}} z - 1\right)$ & $z 
< c_{1}$ \\
  $\hat{p}^{q}(c_{N_{f}})\left(1 - \frac{\hat{p}^{q}(c_{N_{f}})}{2(1 - 
C_{N_{f}})} z\right)$ & $z > c_{N_{f}+1}$
  \end{tabular}
  \right\}.
\end{align}
In this study, we also restrict consideration to regular quantiles $C_{i} 
\equiv i / (N_{f} + 1)$, though \qp\ supports arbitrary quantile spacing.

The quantile parametrization (the namesake of the \qp\ code) is expected to be 
an efficient approximation for \pz s because it allocates storage evenly in the 
space of probability density.
In contrast, the histogram format stores information evenly spaced in redshift, 
and the samples format stores information randomly in probability density.
Depending on the native \pz\ output format, converting to the quantile format 
may require $N_{f}$ numerical optimizations.
We accelerate these optimizations by initializing at rough, approximate 
quantiles based on CDF evaluations on a grid.

\subsection{Comparison Metrics}
\label{sec:metric}

We aim to probe how closely \pz s reconstructed from a limited set of stored 
parameters approximate the original, high-resolution representation 
$\hat{p}^{r}(z)$ of the reference catalog.
This is done without knowledge of a galaxy's true redshift; there is, in fact, 
no notion of a true redshift in our analysis.
(For a demonstration of how one might approach the distinct problem of 
evaluating the accuracy of a \pz\ relative to a true redshift, see 
\citet{polsterer_uncertain_2016}, Schmidt et al.\ in preparation.)

We consider as a metric the loss of information incurred when using an 
approximation of the PDF $\hat{P}(z)$ instead of the reference PDF $P(z)$, 
given by the Kullback-Leibler divergence (KLD), which is defined as
\begin{align}
  \label{eq:kld}
  \mathrm{KLD}[\hat{P}(z) | P(z)] &= \int_{-\infty}^{\infty}\ P(z)\ 
\log\left[\frac{P(z)}{\hat{P}(z)}\right]\ dz,
\end{align}
where $\log$ is the natural logarithm throughout this paper unless otherwise 
indicated, such that the KLD is measured in nats (base $e$ digits, analogous to 
base 2 bits).
Because there is in general no closed-form expression for the KLD, we calculate 
the discrete KLD
\begin{align}
  \label{eq:kld_approx}
  \mathrm{KLD}[\hat{P}(z) | P(z)] &\approx 
\delta_{ff}\sum_{z=z_{1}}^{z_{N_{ff}}}\ P(z)\ 
\log\left[\frac{P(z)}{\hat{P}(z)}\right]
\end{align}
using evaluations of the PDF under each format on a very fine, regular grid 
$(z_{1}, z_{2}, \dots, z_{N_{ff}-1}, z_{N_{ff}})$ with resolution $\delta_{ff} 
= (z_{N_{ff}} -z_{1}) / N_{ff}$ for $N_{ff} \gg N_{f}$.

The most important feature of the KLD is its asymmetry: it is not a distance, 
like the root mean square error, that is the same from $P(z)$ to $\hat{P}(z)$ 
as it is from $\hat{P}(z)$ to $P(z)$.
It is a \textit{divergence} of the information lost when using $\hat{P}(z)$ to 
approximate $P(z)$.
The KLD requires that both functions $P(z)$ and $\hat{P}(z)$ be PDFs (always 
positive semidefinite and integrating to unity); this may need to be explicitly 
enforced for some approximation formats.
The KLD is always positive, and a smaller value indicates better agreement 
between the approximate representation $\hat{p}^{\mathcal{F}}(z)$ and the 
reference representation $p^{r}(z)$.
In the Appendix, we review the properties of the KLD and establish some 
intuition for it.

Additionally, we consider the percent error
\begin{align}
  \label{eq:percent_error}
  \Delta_{m}[\hat{P} | P] &= \frac{M_{m}[P] - 
M_{m}[\hat{P}]}{M_{m}[P]}\times100\%
\end{align}
of the $m^{\mathrm{th}}$ moment
\begin{align}
  \label{eq:moment}
  M_{m}[P] &= \int_{-\infty}^{\infty} z^{m}\ P(z)\ dz\ \approx\  
\delta_{ff}\sum_{z=z_{1}}^{z_{N_{ff}}}\ z^{m}\ P(z)
\end{align}
of a PDF.
We note that $M_{0}[P]=1$ for all properly normalized PDFs, $M_{1}[P]=\bar{z}$ 
is the \textit{mean}, $M_{2}[P]$ is the \textit{variance}, and $M_{3}[P]$ is 
the \textit{skewness}.
Though the first few moments are not in general sufficient to characterize a 
highly structured PDF, they are included in this analysis because they can 
prove useful in setting ballpark estimates of the influence of different 
systematics in various science cases.

The metrics considered in this paper are a subset of those included in the \qp\ 
package and are specific to this investigation; different metrics may be more 
appropriate for other science use cases and indeed could lead to different 
format preferences.

\subsubsection{Individual \pz\ metrics}
\label{sec:individual_metric}

Some science applications rely on the recovery of individual galaxy \pz s that, 
for example, may be used as the basis for finding 
\citep{radovich_searching_2017} or constraining the masses of 
\citep{applegate_weighing_2014} galaxy clusters.
For this purpose, we calculate the KLD of each individual \pz\ in our catalogs 
and then characterize the distribution of KLD values (which is itself a PDF) by 
its $m=1,\ 2,\ 3$ moments.
We also calculate the percent error on the $m=1,\ 2,\ 3$ moments of each \pz\ 
under all parametrizations for both ensembles.
It is natural to expect the KLD distribution moments and moment percent errors 
to decrease with increasing $N_{f}$, indicating convergence toward Gaussian 
errors as the approximation improves.
We use these aggregate statistics to quantify the fidelity of individual \pz\ 
approximations for each dataset as a function of parametrization.

\subsubsection{Stacked $\hat{n}(z)$ estimator}
\label{sec:stacked_metric}

In addition to considering how the choice of storage parametrization affects 
the recovery of individual \pz s, we also demonstrate how one might use 
\texttt{qp} to choose the best parametrization for a particular science case.
We encourage \texttt{qp} users to develop a metric around their own \pz\ use 
cases, as the optimal parametrization may not be shared among all science 
applications of \pz s.

In cosmology, \pz s have thus far been used almost exclusively to estimate the 
redshift distribution function $n(z)$ necessary for calculating the correlation 
functions used by many cosmological probes \citep{clampitt_galaxygalaxy_2017, 
hildebrandt_kids-450:_2017}.
The most common way to estimate the redshift distribution function for a sample 
of $N_{g}$ galaxies is to average the \pz s according to
\begin{align}
  \label{eq:nz}
  \hat{n}(z) &\equiv \frac{1}{N_{g}}\ \sum_{k=1}^{N_{g}}\ \hat{p}_{k}(z),
\end{align}
a procedure producing what we call the stacked estimator $\hat{n}(z)$ of the 
redshift distribution function \citep{harnois-deraps_kids-450:_2017, 
hoyle_dark_2017}.\footnote{
Equation~\ref{eq:nz} is sometimes modified by weights specific to each galaxy 
based on the relative prevalence of galaxies with similar photometry in a 
reference population \citep{sheldon_photometric_2012, 
troster_cross-correlation_2017}
}
Note that to avoid introducing any preferred treatment between formats, 
Equation~\ref{eq:nz} uses the representation $\hat{p}(z)$ reconstructed from 
the stored parameters even when a given format may have a more efficient or 
principled way to obtain such an ensemble estimator directly from the stored 
parameters, e.g. adding the histogram components or combining all samples.
While we do not recommend the approach of Equation~\ref{eq:nz} to estimate the 
redshift distribution (see \citet{choi_cfhtlens_2016} for justification and 
Malz et al., in preparation for alternative methods), we use it here to 
generically demonstrate how one would optimize the choice of \pz\ 
parametrization in the context of a familiar science application.

As the stacked estimator is normalized so that it, too, is a PDF, the KLD 
\textit{from} the approximation (the stacked estimator of a catalog of 
evaluations of reconstructed \pz s) \textit{to} the original (the stacked 
estimator of a catalog of evaluations of the high-resolution reference \pz s) 
serves as a metric for a specific science use case of \pz s.
Because the accuracy of lower-order moments of the redshift distribution 
function dominates the weak lensing error budget, we also compare the percent 
error on the $m=1,\ 2,\ 3$ moments of $\hat{n}(z)$.
However, this information may be less relevant due to the broad range of 
redshifts and small number of galaxies considered in each instantiation.
Furthermore, we note that the dominance of the first few moments of 
$\hat{n}(z)$ may not always hold true as the methodology of \pz\ usage in 
cosmology evolves.

\section{Photo-z Test Data}
\label{sec:data}

With the expectation that the optimal parametrization for approximating \pz s 
may differ according to the properties of the original photometric data, we 
demonstrate a procedure for vetting \pz\ parametrizations on a pair of mock 
datasets, each intended to be realistic predictions of subsets of the 
anticipated LSST \pz s.
All \pz s are fit to simulated LSST 10-year $ugrizy$ apparent magnitudes and 
errors \citep{ivezic_lsst:_2008} using the publicly available Bayesian 
Photometric Redshift (BPZ) code \citep{benitez_bayesian_2000}, which employs 
fitting to a library of spectral energy distribution (SED) templates.
The choice of \pz\ estimation method, however, is not relevant to this study; 
so long as the mock \pz s are \textit{realistically complex}, meaning they take 
shapes with features comparable to those we expect to see in \pz s from real 
datasets with similar photometric properties, it does not matter whether the 
\pz s produced by BPZ are accurate redshift posteriors.
We seek only to optimize the fidelity of the stored \pz\ relative to the 
original \pz\ from a representative \pz\ fitting code.
\citep[See][Schmidt et al.\ in preparation for other work comparing the 
accuracy of \pz s produced by different methods.]{tanaka_photometric_2018, 
de_jong_third_2017, amaro_metaphor:_2016}
As BPZ is a widely used and well established method, we assume that the \pz s 
produced by it are of representative complexity.
The default format of BPZ is a $N_{ff}>200$ gridded parametrization with 
resolution exceeding the available storage for an LSST-like survey.
Because we believe that each galaxy has an underlying redshift interim 
posterior probability density that is a continuous function, to which the 
output of BPZ is itself a high-resolution approximation in the form of 
evaluations on a grid, we fit each gridded \pz\ with a Gaussian mixture model 
that we designate as the reference representation $p^{r}(z)$ for our tests.
The number of components of the mixture model is rounded up from the 
$99^{\mathrm{th}}$ percentile of the modality distribution of the \pz\ catalog 
in question.

\begin{figure*}
  \begin{center}
    \includegraphics[width=\columnwidth]{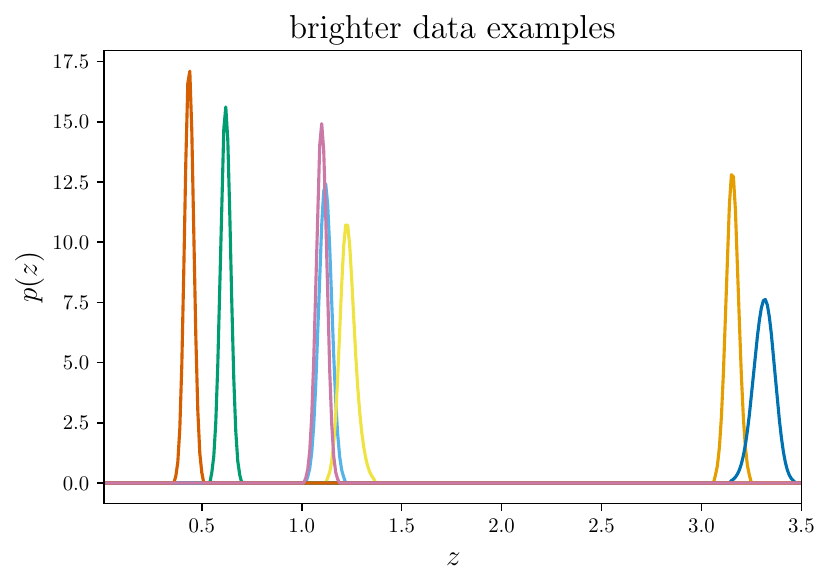}
    \includegraphics[width=\columnwidth]{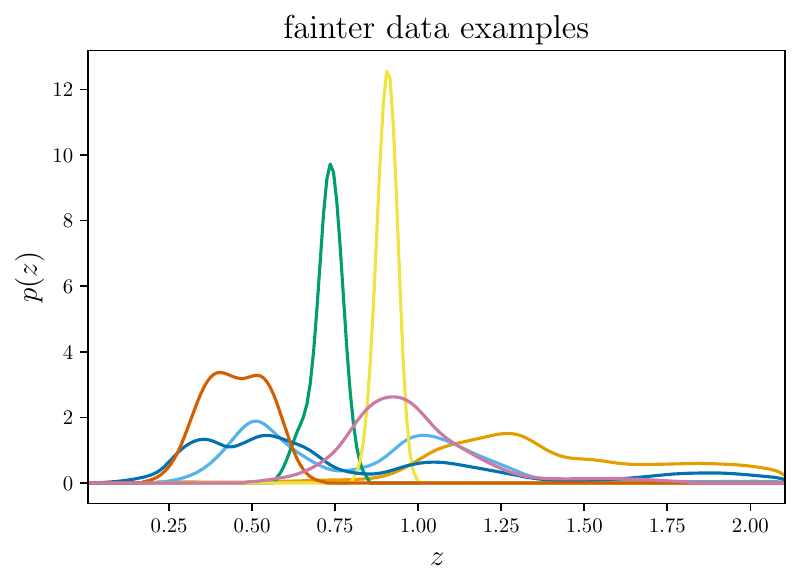}
    \caption{
    Example \pz s from the two mock LSST datasets.
    Left: The \mgdata mock photometry yields largely narrow, unimodal \pz s.
    Right: The \ssdata mock photometry contains a higher proportion of broad 
and/or multimodal \pz s.
    \label{fig:example_pzs}}
  \end{center}
\end{figure*}

\subsection{\Mgdata data mock catalog}
\label{sec:graham}

Our first dataset is an $N_{g} = 10^{5}$ object subset of the 
\citet{graham_photometric_2018} simulated galaxy catalog used for LSST 
photometric redshift experiments.
The data builds on the Millennium simulation of large-scale structure 
\citep{springel_simulations_2005}, the galaxy formation models of 
\citet{gonzalez-perez_how_2014}, and the lightcone construction techniques of 
\citet{merson_lightcone_2013}.
We use the software of \citet{connolly_end--end_2014} to derive observed 
apparent $ugrizy$ magnitudes from the true apparent magnitudes and 10-year LSST 
observational errors of \citet{ivezic_lsst:_2008} before imposing cuts in true 
redshifts $z<3.5$ and observed apparent magnitudes fainter than the predicted 
10-year limiting magnitudes in each filter ($u<26.1$, $g<27.4$, $r<27.5$, 
$i<25$, $z<26.1$, and $y<24.9$) to realistically simulate non-detections.

The \pz\ estimates for this simulated catalog use BPZ templates based on the 
VIsible MultiObject Spectrograph Very Large Telescope Deep Survey set of 
spectra \citep{fevre_vimos_2005}, as in \citet{ilbert_accurate_2006}.
This catalog also uses the default parameter settings for BPZ with the two 
additions of a photometric redshift maximum of 3.5 and an $i$-band magnitude 
prior.
The \pz s from BPZ are in the form of $N_{ff} = 351$ evaluations of the 
probability density on a regular grid of redshifts $0.01 < z < 3.51$, a 
subsample of which are shown in the left panel of Figure~\ref{fig:example_pzs}.
As the figure shows, the \pz s from this dataset tend to be unimodal and 
sharply peaked, as if coming from brighter photometric data due to the 
conservative cuts in photometric magnitudes of this dataset.
The brighter catalog reference \pz s are three-component Gaussian mixtures fit 
to this data.

\subsection{\Ssdata data mock catalog}
\label{sec:schmidt}

Our second dataset is an independent simulation of the expected LSST galaxy 
sample, the Buzzard-highres-v1.0 mock galaxy catalog of deRose et al.\ in 
preparation of galaxies with SEDs drawn from an empirical library of 
$\sim5\times10^{5}$ SEDs from the Sloan Digital Sky Survey (SDSS).
Given an SED, redshift, and absolute $r$-band magnitude for each galaxy, the 
$ugrizy$ magnitudes are derived from the aforementioned 10-year LSST errors.
The catalog contains $N_{g} \approx 10^{5}$ galaxies $z<2.105$ to a depth of 
$i<26.9$, 1.5 magnitudes deeper than the expected LSST gold sample of galaxies 
\citep{ScienceBook}, that will have $S/N \gtrsim 30$ in multiple bands.

We use a custom BPZ prior using a subset of the Buzzard-highres-v1.0 catalog 
and a spanning template set via a simple k-means clustering algorithm based on 
$100$ of the SDSS SEDs used to create the Buzzard catalog.
BPZ produces \pz s in the format of probability density evaluations on a 
regular grid of $N_{ff}=211$ redshifts $0.005\leq z\leq2.105$, a subsample of 
which are plotted in the right panel of Figure~\ref{fig:example_pzs}.
The exceptional depth and known degeneracies (e.~g.~the Lyman/Balmer break 
degeneracy) lead us to expect the presence of multimodal \pz s observed in the 
figure.
The fainter catalog reference \pz s are five-component Gaussian mixtures fit to 
this data.

\section{Results \& Discussion}
\label{sec:results}

We evaluate the metrics of Section~\ref{sec:metric} on 10 random instantiations 
of catalogs of $N_{g}=100$ galaxies drawn randomly from each of the datasets 
discussed in Section~\ref{sec:data} and with each of $N_{f}=3,\ 10,\ 30,\ 100$ 
stored parameters for the three formats of Section~\ref{sec:approx}.
Our analysis characterizes the results using the median and interquartile range 
(IQR), with lower and upper bounds representing the $25^{\mathrm{th}}$ and 
$75^{\mathrm{th}}$ percentiles, as a model-independent error characterization 
appropriate for our small number of mock catalogs.
We then illustrate how our results could be used to choose an appropriate 
parametrization for each dataset given constraints on the distribution of KLDs 
or moment percent errors of individual \pz s, the KLD or moment percent error 
of a science metric ($\hat{n}(z)$ in this case), or the available storage 
capacity.

\subsection{Individual \pz s}
\label{sec:individual_results}

\begin{figure}
  \begin{center}
    \includegraphics[width=\columnwidth]{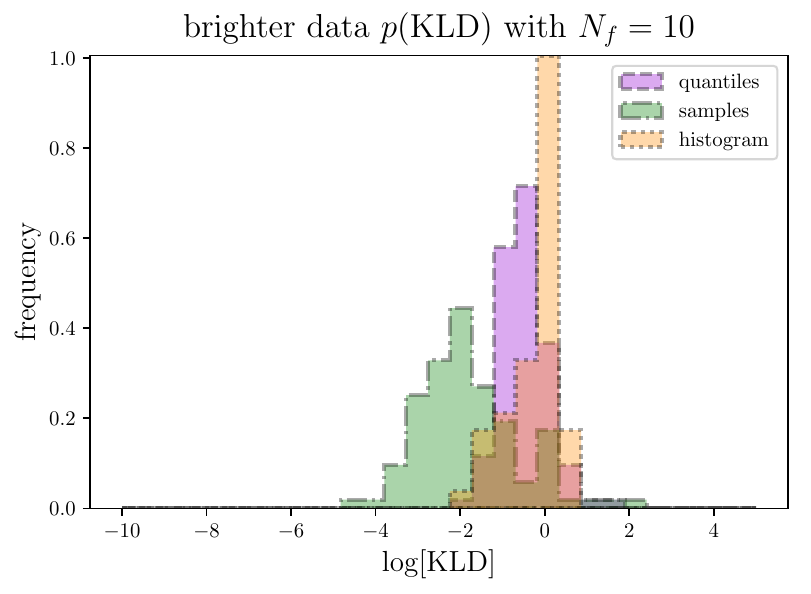}
    \caption{The distribution of log-KLD values for $N_{g}=100$ \pz s from the 
\mgdata dataset with $N_{f}=10$ over the quantiles (purple with dashed border), 
samples (green with dash-dotted border), and histogram (orange with dotted 
border) formats.
    In this instantiation, the samples format has a lower median KLD than the 
quantiles format, which has a lower median KLD than the piecewise constant 
format.
    Note that the distributions are over log-KLD, so the ordering of the 
formats by the breadth of the log-KLD distribution is the same as the order by 
the median.
    \label{fig:individual}}
  \end{center}
\end{figure}

We compare our three formats on the basis of the distributions of the KLD 
calculated for every \pz\ in the two datasets.
An example of an individual \pz\ KLD distribution for the \mgdata dataset with 
$N_{f}=10$ is shown in Figure~\ref{fig:individual}.

\begin{figure*}
  \begin{center}
    \includegraphics[width=\columnwidth]{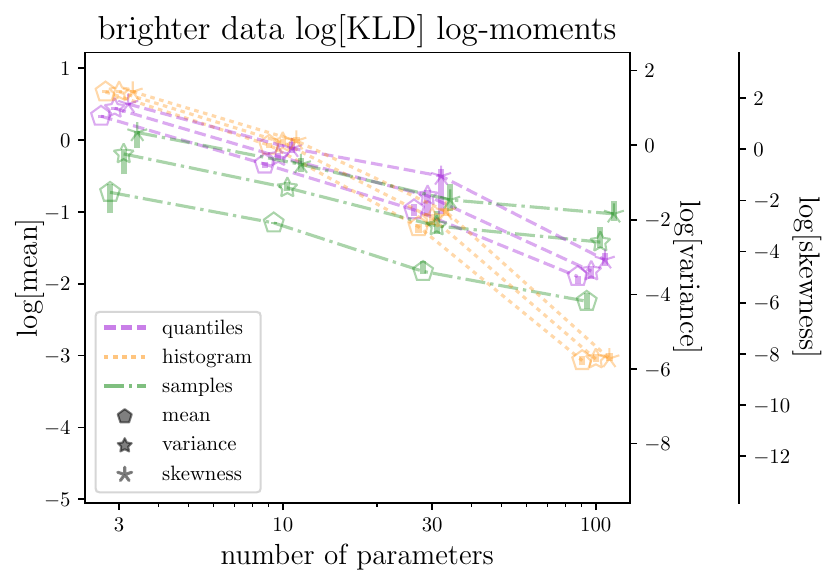}
    \includegraphics[width=\columnwidth]{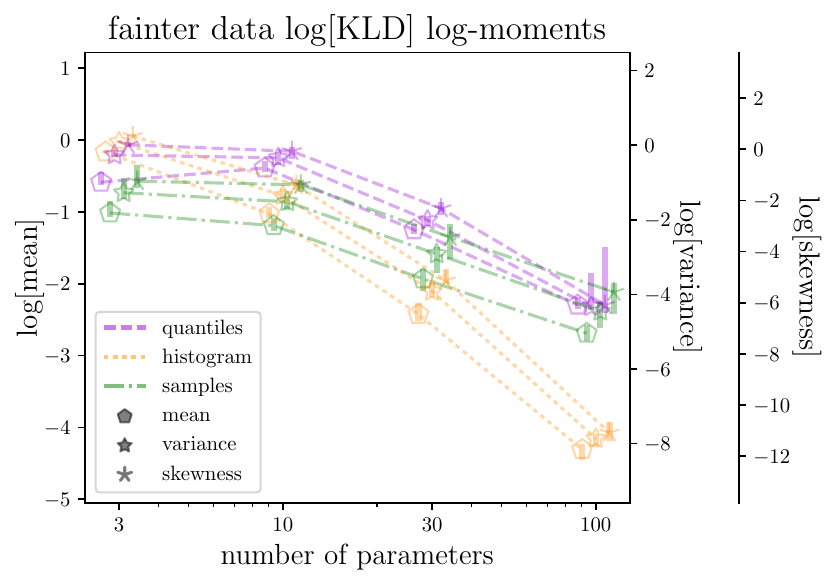}
    \caption{
    The medians of the mean ($\pentagon$), variance ($\bigstar$), and skewness 
($\APLstar$) of the log-KLD distributions for each dataset as a function of the 
number $N_{f}$ of stored parameters for the quantile (purple dashed line), 
samples (green dash-dotted line), and histogram (orange dotted line) formats 
with IQR error bars based on 10 instantiations of 100 galaxies, which are 
offset about $N_{f}$ to improve readability.
    Left panel: The moments of the distribution of individual \pz\ log-KLD 
values of the \mgdata mock catalog are minimized when they are stored as 
samples, except at large $N_{f}$.
    Right panel: The \ssdata mock catalog achieves equivalence of the formats 
in the moments of the log-KLD distributions at a much lower $N_{f}$, ultimately 
showing the histogram format minimizes the moments at all but the smallest 
$N_{f}$.
    \label{fig:kld_moments}}
  \end{center}
\end{figure*}

To distill what is observed in the ten instantiations of plots like 
Figure~\ref{fig:individual} for both datasets and all parametrizations, we 
compare the first three moments of the distributions of metric values for the 
distribution of the KLDs of individual \pz s under each parametrization, 
summarized in Figure~\ref{fig:kld_moments}.
While it is obvious that one would like the mean of the KLD distribution to be 
low, interpretation of higher-order moments is less clear.
In a science application that is robust to \pz\ outliers, a parametrization 
with a high variance or skewness may be acceptable, whereas another science 
application that simply requires a well-characterized error distribution may 
tolerate a higher mean in exchange for lower variance and skewness.
To meaningfully interpret the KLDs of individual \pz s, it will be necessary 
for those using \pz s in their science to calculate the requirements on the 
acceptable degree of information loss.

As expected, the behavior of the log-moments of the log-KLD distribution is 
highly correlated for a given format and number of parameters, for both 
datasets.
The \mgdata dataset has slightly higher log-KLD log-moments than the \ssdata 
dataset at all $N_{f}$ and across all formats, meaning information loss is 
enhanced for more strongly featured data; this observation is not surprising 
because the narrow, unimodal \pz s of the \mgdata dataset have long tails of 
very low probability that are emphasized by the KLD.
Both datasets exhibit decreasing moments for the quantile and samples formats 
as $N_{f}$ increases, though the marginal improvement at high $N_{f}$ is 
greater for the \ssdata dataset.

The log-KLD log-moments are higher for quantiles than for samples in both 
datasets, except at $N_{f}=100$ for the \mgdata dataset.
The histogram format's log-KLD log-moments are higher than those of other 
formats at the lowest $N_{f}$ and steadily decrease in a manner similar to the 
other formats, except at the highest $N_{f}$ values where the histogram 
format's log-KLD log-moments decrease much more quickly.
Neither of these results are unexpected because of the KLD's sensitivity to the 
tails; our choise of PDF reconstruction method for the quantile format is most 
susceptible to error in the tails of the PDF, and only the histogram format 
preserves information uniformly at all redshifts rather than at all 
probabilities.

\begin{figure*}
  \begin{center}
    \includegraphics[width=\columnwidth]{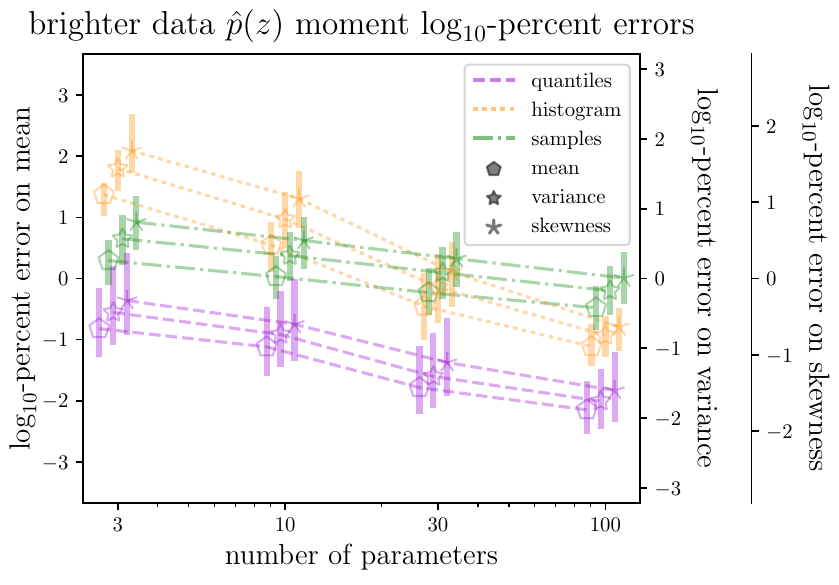}
    \includegraphics[width=\columnwidth]{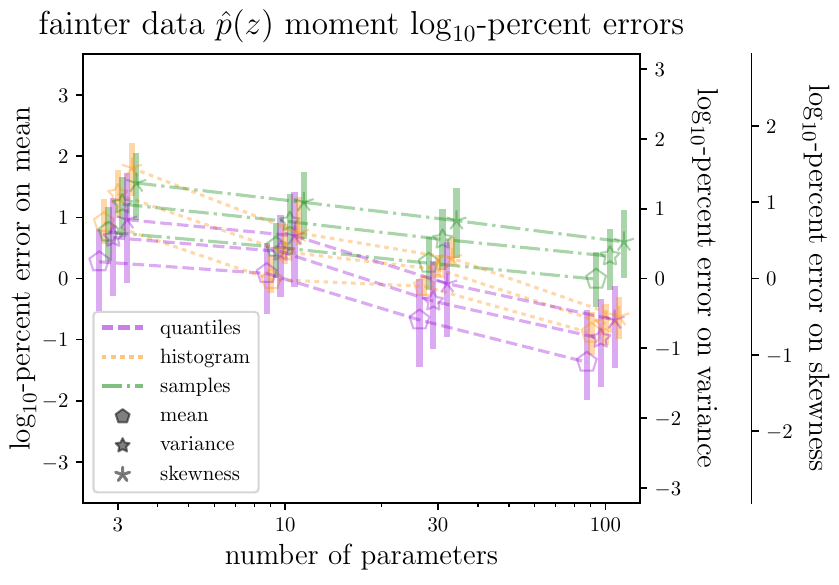}
    \caption{
    The median log$_{10}$-percent errors on the mean ($\pentagon$), variance 
($\bigstar$), and skewness ($\APLstar$) of the \pz s for each dataset as a 
function of the number $N_{f}$ of stored parameters per \pz\ for the quantile 
(purple dashed line), samples (green dash-dotted line), and histogram (orange 
dotted line) formats with IQR error bars based on 10 instantiations of 100 
galaxies, where the log$_{10}$-percent errors and their IQRs are offset about 
$N_{f}$ to improve readability.
    Left panel: The \mgdata \pz\ ensemble's moment percent errors are minimized 
by the quantile format at all $N_{f}$.
    Right panel: The \ssdata \pz\ ensemble's moment percent errors are high for 
all formats at low $N_{f}$ but distinct at high $N_{f}$, with the quantile 
format overall outperforming the samples and histogram formats.
    \label{fig:pz_moment_errs}}
  \end{center}
\end{figure*}

In Figure~\ref{fig:pz_moment_errs}, we also examine the percent error on the 
first three moments of the \pz s under each approximation, using the base-10 
log for interpretability.

Note that the error bars here are larger because they include the effects of 
variation between individual galaxies, whereas all other plots of moments have 
error bars that represent only the variation due to catalog instantiations.

Though the log$_{10}$-percent error of the moments of individual \pz s also 
exhibits significant correlation between the moments for a given 
parametrization, the behavior is otherwise markedly different from that of the 
log-moments of the \pz\ ensemble's log-KLD distribution.
The percent errors of the moments of the approximate \pz s are overall lower in 
the \mgdata dataset than those of the \ssdata dataset over the same range of 
number of stored parameters; this is expected because there is simply less 
information to capture in the \mgdata dataset.
The unwaveringly linear marginal improvement in the log-percent error of the 
per-galaxy \pz\ moments with increasing log-$N_{f}$ may reflect the fact that 
samples are taken randomly in the space of probability and thus unaffected by 
the interpolation artifacts of the quantile format and the interplay between 
the scales of features and bins of the histogram format.

For the \mgdata dataset, the quantile format is the only one that consistently 
achieves sub-percent errors in \pz\ moments even at low $N_{f}$.
Furthermore, for the \mgdata dataset, the quantile format minimizes the percent 
error at all $N_{f}$, whereas the samples format outperforms the histogram 
format at low $N_{f}$ before the histogram format overtakes it at high $N_{f}$.
Again, this behavior is expected of the narrow, unimodal \pz s of the \mgdata 
dataset because large histogram bins are ineffective at capturing small-scale 
structure and including more samples does not significantly improve 
preservation of such features.

The qualitative behavior of the moment percent error of all formats is the same 
for the \ssdata dataset as that of the \mgdata dataset at $N_{f}=3$.
In the \ssdata dataset, the inclusion of $N_{f}=30$ parameters decreases the 
moment percent error of the histogram format more significantly than the 
quantile or samples formats, to the point that the histogram and quantile 
formats have comparable moment percent errors.
At higher $N_{f}$ in the \ssdata dataset, the quantile and histogram formats 
continue to improve faster than the samples format, with the percent errors on 
the \pz\ moments being consistently lower for the quantile format than for the 
histogram format.
The broad, multimodal \pz s of the \ssdata dataset enable achievement of 
sub-percent accuracy in the moments only with $N_{f}\geq30$ under the quantile 
format and $N_{f}=100$ with the histogram format.

\subsection{Stacked $\hat{n}(z)$ estimator}
\label{sec:stacked_results}

\begin{figure}
  \begin{center}
    \includegraphics[width=\columnwidth]{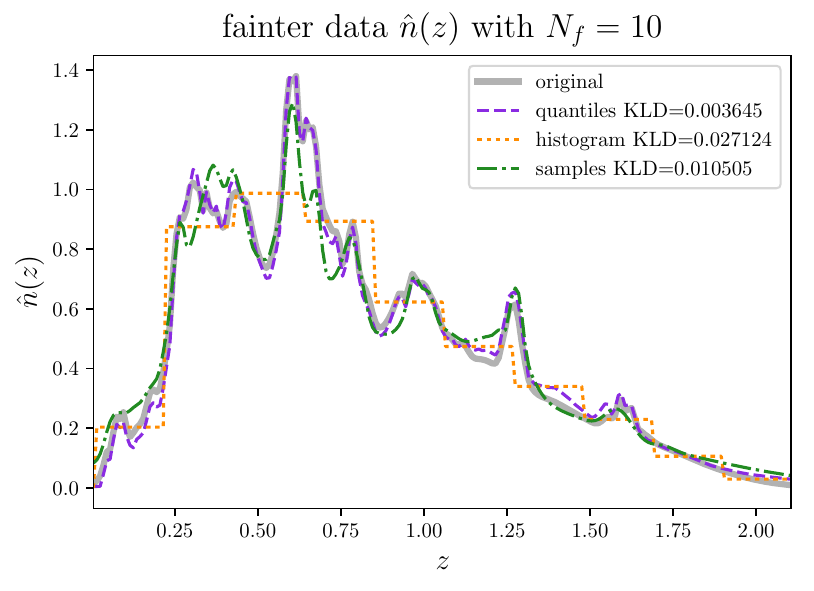}
    \caption{An example of the stacked estimator of the redshift distribution, 
for a subsample of $N_{g}=100$ galaxies drawn from the \ssdata data mock 
catalog and with $N_{f}=10$ parameters used for each \pz; the small-scale 
features are due to the small number of galaxies in the sample.
    The most striking characteristic of $\hat{n}(z)$ of a small number of 
galaxies with a relatively small number of parameters is the coarseness of the 
histogram format (orange dotted line) relative to the quantile (purple dashed 
line) and samples (green dash-dotted line) formats, both of which are fairly 
close to $\hat{n}(z)$ derived from evaluating the original, high-resolution \pz 
s (thick gray line).
    \label{fig:stacked}}
  \end{center}
\end{figure}

Figure~\ref{fig:stacked} shows an example of $\hat{n}(z)$ of \pz s 
reconstructed from just $N_{f}=10$ parameters under each of our three 
approximation formats, evaluated on the same fine grid as the input \pz s.
The strong features in the curve are due to the small sample size of 
$N_{g}=100$ galaxies.
As expected, the stacked histogram is quite coarse because of the step function 
interpolation, while the stacked estimator of the redshift distribution based 
on \pz\ representations that are interpolations of stored samples and quantiles 
are much closer to the stacked estimator of the original, high-resolution \pz s.
The KLD for each format is also included in the plot; in this instance, the KLD 
is lowest for the quantile format and highest for the histogram format.

\begin{figure*}
  \begin{center}
    \includegraphics[width=\columnwidth]{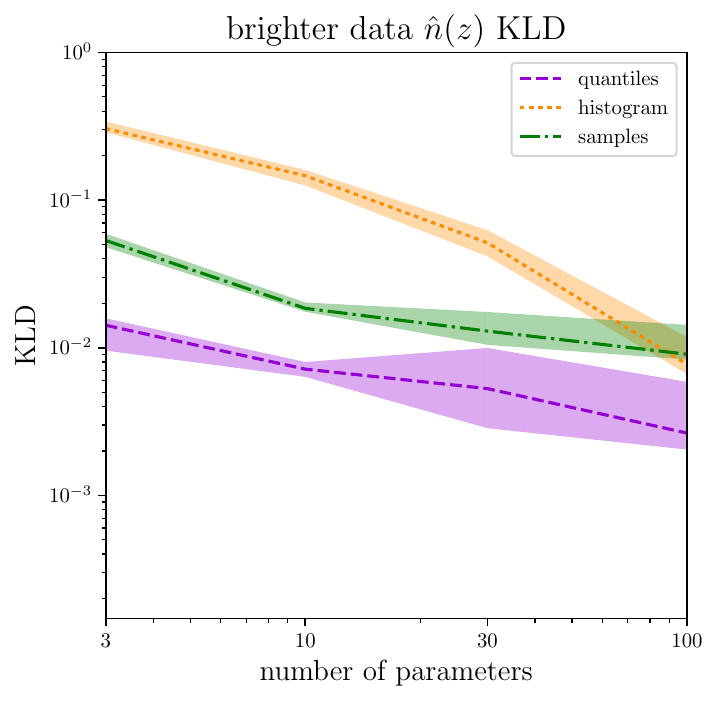}    
\includegraphics[width=\columnwidth]{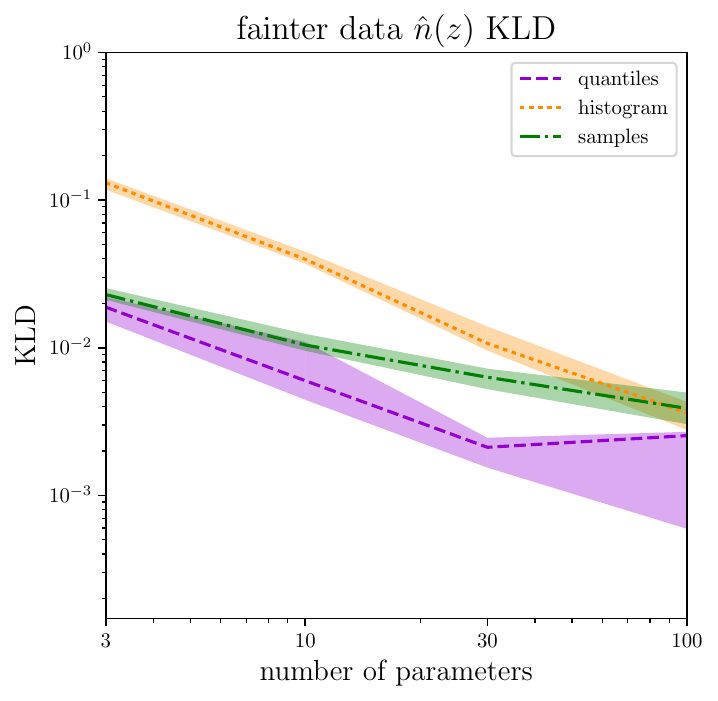}
    \caption{
    The KLD between $\hat{n}(z)$ derived from approximate \pz\ representations 
and $\hat{n}(z)$ derived from the original, high-resolution \pz s, as a 
function of number $N_{f}$ of stored parameters, for the quantiles (purple 
dashed line), samples (green dash-dotted line), and histogram (orange dotted 
line) formats.
    Shaded regions indicate the IQR errors derived from 10 subsamples of 100 
galaxies and lines indicate the median of the distribution.
    Left panel: The \mgdata \pz\ catalog's KLD of $\hat{n}(z)$ is minimized by 
the quantile format at all $N_{f}$.
    Right panel: The \ssdata \pz\ catalog's KLD of $\hat{n}(z)$ is minimized by 
the quantile format at all $N_{f}$, though the samples format's performance is 
comparable.
    \label{fig:kld}}
  \end{center}
\end{figure*}

Again, due to the variation between $N_{g}=100$ galaxy subsamples, we repeat 
$10$ times the procedure that produced Figure~\ref{fig:stacked} to generate a 
distribution over the KLD of the stacked estimator of the redshift distribution 
for each format and dataset.
The $\hat{n}(z)$ KLD values for each parametrization on both mock datasets are 
collected and plotted in Figure~\ref{fig:kld}, with error regions based on the 
IQR of the 10 instantiations.

Figure~\ref{fig:kld} shows that the two datasets clearly share some features:
\begin{enumerate}
\item As expected, the KLD drops as the number of stored parameters increases, 
for all formats.
\item The quantile format minimizes the KLD at all numbers of stored parameters 
considered.
\item The histogram format leads to substantial loss of information relative to 
the other formats except at large numbers of stored parameters where it is on 
par with the samples format.
\end{enumerate}

However, there are also ways in which the behavior of the KLD on $\hat{n}(z)$ 
differs due to the data quality's significant impact on this metric:
\begin{enumerate}
\item The \ssdata dataset in general achieves lower KLD values than the \mgdata 
dataset, likely a consequence of the strong features present in $\hat{n}(z)$ 
for the \mgdata dataset in our subsamples of 100 galaxies.
\item The rate of marginal improvement in the KLD of $\hat{n}(z)$ with 
increasing $N_{f} $ is lower in the \mgdata dataset than the \ssdata dataset; 
in other words, saving more parameters may have a greater marginal benefit for 
the \ssdata dataset than for the \mgdata dataset.
\item The $\hat{n}(z)$ KLD of the samples format is not substantially higher 
than that of the quantile format in the \ssdata dataset but is for the \mgdata 
dataset, which may reflect the subjectivity of the reconstruction scheme used 
for those two formats.
\end{enumerate}

We also address the relative, marginal, and absolute performance and 
consistency thereof of the KLD on $\hat{n}(z)$ for each parametrization as a 
function of format and $N_{f}$ for each dataset.
To guide this process, we interpret Figure~\ref{fig:kld} in the context of 
constraints on storage allocation imposed by the survey and constraints on the 
acceptable degree of information loss imposed by the science requirements, 
which we anticipate establishing in the future.

A constraint on storage resources corresponds to a vertical line at a given 
$N_{f, \mathrm{lim}}$ in Fig. \ref{fig:kld}; the best format would be the one 
that achieves the lowest KLD at $N_{f, \mathrm{lim}}$.
For example, if $N_{f, \mathrm{lim}}=10$ stored parameters, the quantile format 
would be optimal for the \mgdata dataset because it has the lowest KLD value by 
a large margin compared to other formats.
If the \ssdata dataset were subject to the same constraint, the quantile and 
samples formats could both be good candidates for a storage parametrization, 
with the quantile format opening the possibility of a lower KLD.

A constraint on the acceptable loss of information due to compression and 
reconstruction of \pz s corresponds to a horizontal line at some 
$\mathrm{KLD}_{\mathrm{lim}}$ in Figure~\ref{fig:kld}; the best parametrization 
would correspond to the format that achieves $\mathrm{KLD}_{\mathrm{lim}}$ at 
the lowest $N_{f}$.
For example, if our science requires $\mathrm{KLD}_{\mathrm{lim}}=10^{-2}$ 
nats, the optimal parametrization would be quantiles with $N_{f}=3$ for the 
\mgdata dataset and quantiles with $N_{f}=10$ for the \ssdata dataset.

If there is some flexibility in the allocation of storage for \pz s, as is the 
case for LSST, it is valuable to examine the asymptotic behavior of the KLD as 
a function of the number of stored parameters for each format considered.
It may be possible to request additional storage resources for the survey's \pz 
s if the KLD is significantly reduced with a slightly larger $N_{f}$.

\begin{figure*}
  \begin{center}
    \includegraphics[width=\columnwidth]{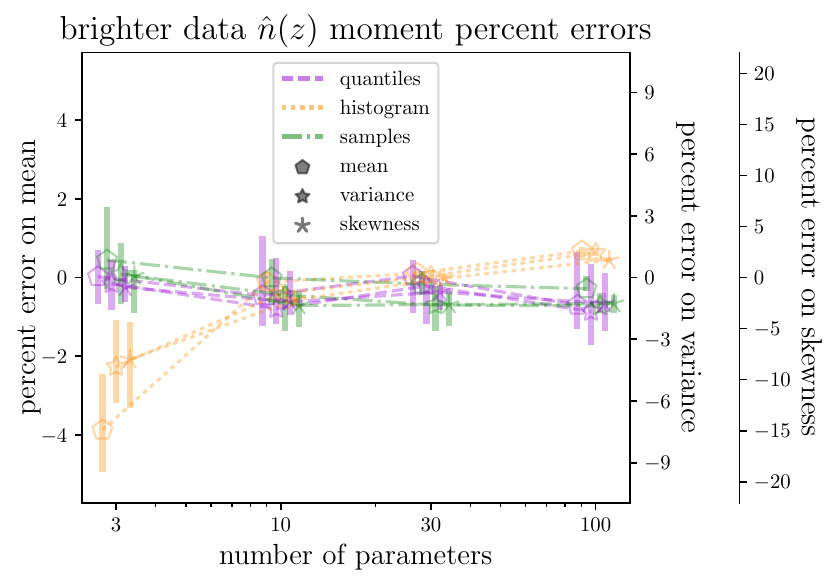}    
\includegraphics[width=\columnwidth]{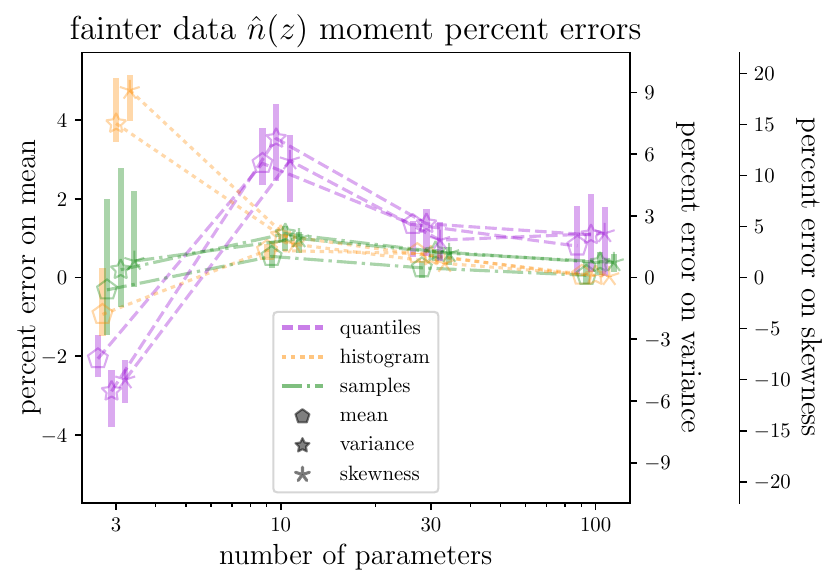}
    \caption{
    The median percent error on the mean ($\pentagon$), variance ($\bigstar$), 
and skewness ($\APLstar$) of the stacked estimator $\hat{n}(z)$ of the redshift 
distribution for each dataset as a function of the number $N_{f}$ of stored 
parameters for the quantile (purple dashed line), samples (green dash-dotted 
line), and histogram (orange dotted line) formats with IQR error bars based on 
10 instantiations of 100 galaxies, where the percent errors and their error 
bars are offset about $N_{f}$ to improve readability.
    Left panel: The \mgdata dataset shows evolution with $N_{f}$ of the 
$\hat{n}(z)$ moment percent errors for the histogram format but none for the 
samples and quantile formats.
    Right panel: The \ssdata dataset shows qualitatively different evolution 
with $N_{f}$ of the $\hat{n}(z)$ moment percent errors for the three formats 
and for each moment.
    \label{fig:nz_moment_errs}}
  \end{center}
\end{figure*}

We also calculate the percent error on the moments of the stacked estimator of 
the redshift distribution, as these may be more useful for understanding error 
propagation in cosmology due to \pz\ storage parametrization than the KLD, for 
which no such infrastructure yet exists.
The percent error on the first three moments of the stacked estimator of the 
redshift distribution function is shown in Figure~\ref{fig:nz_moment_errs}, and 
it is clear that the photometric data quality dominates this metric.

To explain this, we draw attention to Figure~\ref{fig:stacked} and note that, 
though the true distribution of redshifts is similar, the redshift range over 
which they are defined is larger for the \mgdata dataset than the \ssdata 
dataset.

In the \mgdata dataset, the evolution of the $\hat{n}(z)$ moment errors with 
$N_{f}$ differs for the histogram format relative to the samples and quantile 
formats, which exhibit essentially no evolution in excess of the error bars 
between instantiations.
For $N_{f}=3$, the histogram format's moments are grossly underestimated 
because most of the probability density of $\hat{n}^{r}(z)$ falls into the 
lowest redshift bin, and most of the probability density of $\hat{n}^{r}(z)$ is 
above the middle of that bin.
When the bins are too small, at $N_{f}=100$, those at high redshifts have most 
of their probability density below the middle of the bin, leading to slightly 
overestimated moments.
Because the \pz s in the \mgdata dataset are so narrow and unimodal overall, 
the reconstructions of the samples and quantile parametrizations are highly 
accurate where most of the probability density is, even with low $N_{f}$, so 
the reference representation moments are consistently recovered to within 
$<1\%$.

In the \ssdata dataset, the issues are different because the redshift range of 
the original \pz s is smaller and the \pz s themselves are broader.
The samples format has no significant evolution in moment errors with $N_{f}$, 
the histogram format severely overestimates the higher moments at low $N_{f}$, 
and the quantiles format severely underestimates the moments at low $N_{f}$, 
severely overestimates them at intermediate $N_{f}$, and moderately 
overestimates them at high $N_{f}$.
The samples format may suffer from shot noise for broad, multimodal \pz s, but 
the result is just spikier \pz s that produce narrow features in $\hat{n}(z)$ 
that do not significantly affect the moments, explaining the minimal impact on 
the percent errors of the moments of $\hat{n}(z)$.
The histogram format's overestimation of the variance and skewness at low 
$N_{f}$ in the \ssdata dataset is caused by the bulk of the probability density 
of $\hat{n}^{r}(z)$ falling almost evenly into the two low redshift bins with 
far less probability in the highest bin.

As was hinted at in Figure~\ref{fig:kld_moments}, the quantile 
parametrization's \pz\ KLD distribution has large moments, and the KLD is most 
sensitive to a poor approximation of the tails of the distribution.
Both the underestimation of the $\hat{n}(z)$ moments at low $N_{f}$ and the 
overestimation of the $\hat{n}(z)$ moments at intermediate $N_{f}$ are due to 
the choice of a suboptimal reconstruction scheme for quantiles that could 
doubtlessly be improved in the future.
The quantile format's overestimation of the moments even at high $N_{f}$ can be 
explained by the fact that \qp\ does not limit the quantile values to the 
redshift range over which the original \pz s were defined.
A broad \pz\ may thus be reconstructed with probability density outside the 
redshift range of the original \pz s and then truncated and normalized prior to 
calculating the KLD.
Because broad \pz s are more likely to occur at high redshift, this excess 
probability is more likely to be at high redshift, slightly but consistently 
inflating all moments.

\section{Conclusions \& Future Directions}
\label{sec:conclusions}

This work develops a principled, systematic approach to choosing a 
parametrization for storing a catalog of \pz s from a survey of known data 
properties with a goal of balancing the available storage resources against the 
accuracy of the \pz s and science products thereof reconstructed from the 
stored parameters.
We demonstrate the recommended method on two realistic mock datasets 
representative of upcoming \pz\ catalogs and draw the following conclusions:

\begin{itemize}
  \item Some general trends are shared among the datasets we used in our tests, 
but much of the qualitative and quantitative behavior is different.
  The properties of the \pz\ catalog influence the optimal compression scheme.
  \item The parametrization that best approximates individual \pz s may differ 
from the parametrization that optimizes a given science metric.
  The science goals must motivate the metric that guides the choice of 
parametrization.
  \item In our LSST-like examples with metrics motivated by gravitational 
lensing probes of cosmology, we confirm the expectation that regular binning 
and uniform sampling in the space of probability is more effective than regular 
binning in redshift.
  This trend can only be enhanced as the quantile and sample reconstruction 
schemes improve.
\end{itemize}

To be clear, we do not advocate for a one-size-fits-all solution to the problem 
of compressing \pz\ catalogs and emphasize that any decision should be 
motivated by science requirements and account for the absolute, relative, and 
marginal behavior of the formats considered as a function of the number of 
stored parameters.

For the case of LSST, though the histogram format has the strongest presence in 
the \pz\ literature, it exhibits a higher loss of information and moment 
percent error of the reconstructed \pz s, except when a very large number of 
parameters are stored, so we do not recommend its use for LSST's \pz\ catalog.
Given the constraint that LSST will be able to store only $\sim100$ numbers to 
describe the redshift of each galaxy and intends to include the output of 
several \pz\ codes, we can safely say that LSST can store the output of more 
than one \pz\ code without risk of significant loss of information.
Had our results indicated a significant improvement in our metrics for a small 
increase in the number of stored parameters, we would present to 
decision-makers within the collaboration evidence in support of increasing that 
allocation.

Furthermore, though we discussed the previous use of each format in science 
calculations, we do not endorse any format on the basis of existing 
infrastructure for a particular science application.
Rather, we anticipate great advances in the development of analysis techniques 
that best make use of the information in \pz s and encourage the community to 
then choose parametrizations that most effectively serve the needs of those 
intended practices.
Future analyses may also consider options we did not, such as additional 
formats, new metrics, variable $N_{f}$ over the PDF ensemble, irregular spacing 
of shared parameters $\vec{C}$, and improved samples and quantile 
reconstruction procedures.

So that decisions of this kind can be optimized for all future surveys, the 
\qp\ Python package developed for this project is made public on GitHub as a 
tool for use by the broader community.
We invite contributions of formats, metrics, and reconstruction schemes to the 
public GitHub repository.

\subsection*{Appendix}
\label{sec:kld}

We develop some intuition for the Kullback-Leibler Divergence by contrasting it 
with the familiar metric of the root-mean-square error (RMSE)
\begin{align}
  \label{eq:rmse}
  \mathrm{RMSE} &= \sqrt{\int (P(z) - \hat{P}(z))^{2} dz}.
\end{align}
Consider the simple example of a Gaussian $P(z) = \mathcal{N}(\mu_{0}, 
\sigma_{0}^{2})$ being approximated by a Gaussian $\hat{P}(z) = 
\mathcal{N}(\mu, \sigma^{2})$, whose KLD is
\begin{align}
  \label{eq:gaussian}
  \mathrm{KLD} &= 
\frac{1}{2}\left(\log\left[\frac{\sigma^{2}}{\sigma_{0}^{2}}\right] + 
\frac{\sigma_{0}^{2}}{\sigma^{2}} + \frac{(\mu-\mu_{0})^{2}}{\sigma^{2}} - 
1\right)
\end{align}
To get a sense of the units of information, we can calculate the KLD and RMSE 
in some limiting cases.
If $\sigma=\sigma_{0}$ but $\mu=\mu_{0}+1$, we obtain 
$\mathrm{KLD}=\frac{1}{2}$ nat --- if the mean of the approximation is wrong by 
an additive factor of $\sigma$, half a nat of information is lost.
If $\mu=\mu_{0}$ but $\sigma=\sqrt{2\pi}\sigma_{0}$, we find 
$\mathrm{KLD}\approx\frac{1}{2}$ nat --- half a nat of information is also lost 
if the variance of the approximation is off by a multiplicative factor of 
$2\pi$.

We can use the KLD to identify notions of imprecision and inaccuracy.
Intuitively, precision must be related to how close $\sigma$ is to $\sigma_{0}$ 
and accuracy must be related to how close $\mu$ is to $\mu_{0}$.

\begin{figure}
  \begin{center}
    \includegraphics[width=\columnwidth]{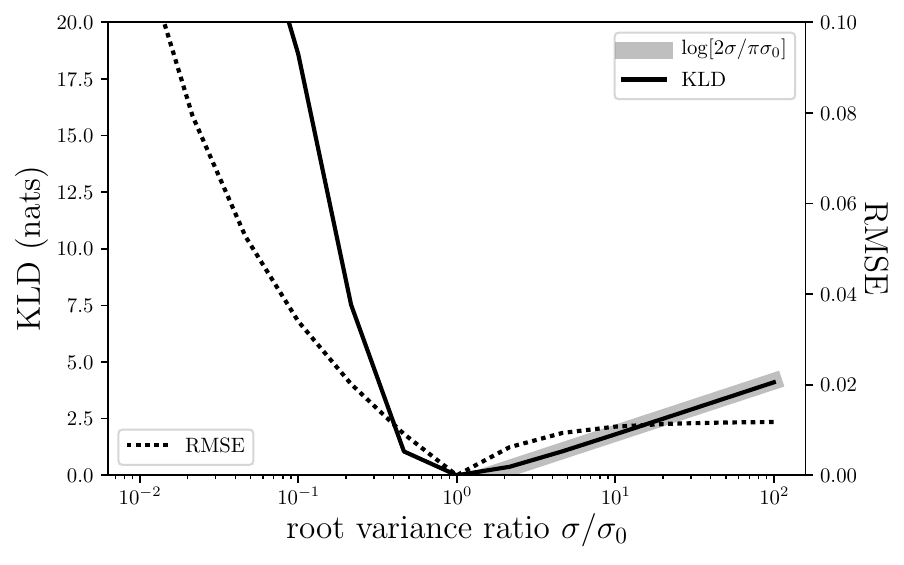}
    \caption{The KLD and RMSE as a function of the root variance ratio $r$ for 
a simple Gaussian example.
    The KLD (solid line) rises sharply at $\sigma<\sigma_{0}$ and is 
proportional to the log of the inverse precision $r$ for $\sigma>\sigma_{0}$, 
behavior that is qualitatively similar to that of the RMSE (dotted line).
    \label{fig:precision}}
  \end{center}
\end{figure}

If $\mu\approx\mu_{0}$, we can say $\mathrm{KLD}\sim\log[r] + \frac{1}{2}r^{-2} 
- \frac{1}{2}$ where $r^{-1}\equiv\frac{\sigma_{0}}{\sigma}$ is a measure of 
\textit{precision}, whose behavior is illustrated in 
Figure~\ref{fig:precision}, alongside that of the RMSE.  We observe that an 
overestimated variance increases the KLD as the log of the square root of the 
ratio of the estimated variance to the true variance.

\begin{figure}
  \begin{center}
    \includegraphics[width=\columnwidth]{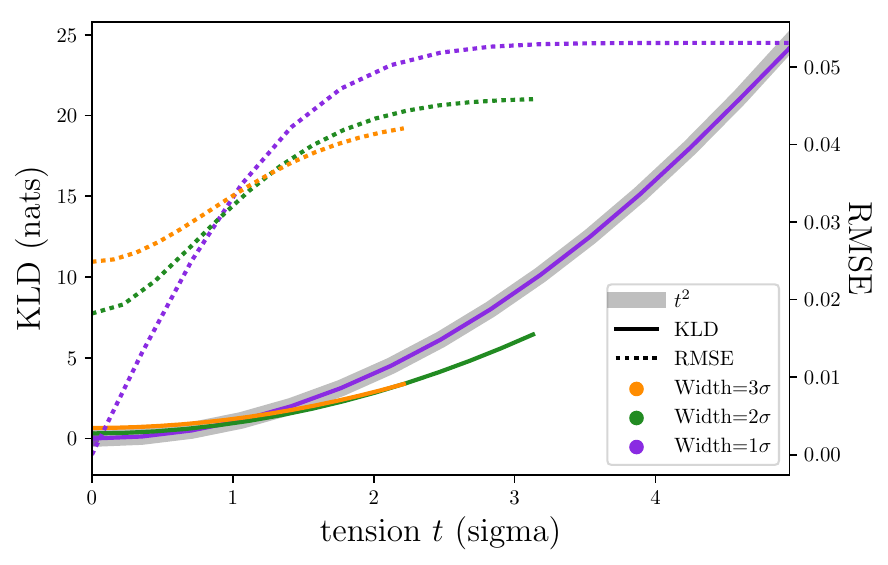}
    \caption{The KLD and RMSE as a function of the tension $t$ for a simple 
Gaussian example.
    The KLD (solid lines) is equal to the square of the tension $t$, with a 
small offset when $r\neq1$, whereas the RMSE (dotted lines) is relatively 
insensitive to tension past a certain point but more sensitive to $r\neq1$.
    \label{fig:tension}}
  \end{center}
\end{figure}

When $\sigma\approx\sigma_{0}$, $\mathrm{KLD}\sim t^{2}$ in terms of the 
\textit{tension} $t\equiv\frac{(\mu-\mu_{0})^{2}}{\sigma^{2}}$, whose 
concordance is illustrated in Figure~\ref{fig:tension}.
There is some limiting tension $t_{\mathrm{lim}}\approx2$ below which the RMSE 
is more sensitive than the KLD and above which the KLD is more sensitive than 
the RMSE.
This behavior hints at the KLD's reputation for sensitivity to the tails of the 
reference PDF.
The notion of tension may be more important for cosmological applications of 
\pz s, indicating the KLD may be a more appropriate metric for coarser 
approximations and the RMSE may be a more appropriate metric for less coarse 
approximations.

\subsection*{Acknowledgments}

This work was incubated at the 2016 LSST-DESC Hack Week hosted by Carnegie 
Mellon University.
AIM is advised by David W. Hogg and was supported by National Science 
Foundation grant AST-1517237.
The work of AIM was also supported by the U.S. Department of Energy, Office of 
Science, Office of Workforce Development for Teachers and Scientists, Office of 
Science Graduate Student Research (SCGSR) program, administered by the Oak 
Ridge Institute for Science and Education for the DOE under contract number 
DE‐SC0014664.
The work of PJM was supported by the U.S. Department of Energy under contract 
number DE-AC02-76SF00515.
SJS was partially supported by the National Science Foundation under grant 
N56981CC.

We would like to thank Chad Schafer, Chris Morrison, Boris Leistedt, Stefano 
Cavuoti, Maciej Bilicki, Johann Cohen-Tanugi, Eric Gawiser, Daniel Gruen, and 
Seth Digel for helpful feedback in the preparation of this paper.
AIM thanks Coryn Bailer-Jones, Eric Ford, and Morgan Fouesneau for providing 
context for this work outside cosmology.

%
%

The DESC acknowledges ongoing support from the Institut National de Physique Nucl\'eaire et de Physique des Particules in France; the Science \& Technology Facilities Council in the United Kingdom; and the Department of Energy, the National Science Foundation, and the LSST Corporation in the United States.
DESC uses resources of the IN2P3 Computing Center (CC-IN2P3--Lyon/Villeurbanne - France) funded by the Centre National de la Recherche Scientifique; the National Energy Research Scientific Computing Center, a DOE Office of Science User Facility supported by the Office of Science of the U.S.\ Department of Energy under Contract No.\ DE-AC02-05CH11231; STFC DiRAC HPC Facilities, funded by UK BIS National E-infrastructure capital grants; and the UK particle physics grid, supported by the GridPP Collaboration.
This work was performed in part under DOE Contract DE-AC02-76SF00515.

Author contributions are listed below. \\
A.I.~Malz: Conceptualization, formal analysis, investigation, methodology, software, visualization, writing -- original draft; initiated project, led development work. \\
P.J.~Marshall: Project administration, software, supervision, writing -- original draft; advised on statistics, and project design and management. \\
J.~DeRose: Resources; produced the photometry for the fainter mock catalog. \\
M.L.~Graham: Resources, writing -- review \& editing; produced the photometry and PDFs for the brighter mock catalog. \\
S.J.~Schmidt: Resources, writing -- review \& editing; produced the PDFs for the fainter mock catalog. \\
R.~Wechsler: Resources; assisted in producing the fainter mock catalog photometry. \\

\software{
jupyter \citep{kluyver_jupyter_2016},
matplotlib \citep{hunter_matplotlib:_2007},
numpy \citep{walt_numpy_2011},
pathos \citep{mckerns_building_2012, mckerns_pathos:_2010},
qp \citep{malz_qp_2017},
scikit-learn \citep{pedregosa_scikit-learn:_2011},
scipy \citep{jones_scipy:_2001}
}

\bibliography{lsstdesc,main}

\end{document}